\documentclass[pra,floatfix,superscriptaddress,twocolumn]{revtex4}
\usepackage{graphicx}

\begin{document}

\title{Strong Dissipation Inhibits Losses and Induces Correlations in Cold Molecular Gases}

\author{N. Syassen}
\author{D.~M. Bauer}
\author{M. Lettner}
\affiliation{Max-Planck-Institut f{\"u}r Quantenoptik, Hans-Kopfermann-Stra{\ss}e 1, 85748 Garching, Germany}
\author{T. Volz}
\altaffiliation{Present address: Institute of Quantum Electronics, ETH-H\"{o}nggerberg, 8093 Z\"{u}rich, Switzerland}
\affiliation{Max-Planck-Institut f{\"u}r Quantenoptik, Hans-Kopfermann-Stra{\ss}e 1, 85748 Garching, Germany}
\author{D. Dietze}
\altaffiliation{Present address: Technische Universit\"{a}t Wien, Institut f\"{u}r Photonik, Gu{\ss}hausstr. 25-29, 1040 Wien, Austria} 
\affiliation{Max-Planck-Institut f{\"u}r Quantenoptik, Hans-Kopfermann-Stra{\ss}e 1, 85748 Garching, Germany}
\author{J.~J. Garc\'{i}a-Ripoll}
\affiliation{Max-Planck-Institut f{\"u}r Quantenoptik, Hans-Kopfermann-Stra{\ss}e 1, 85748 Garching, Germany}
\affiliation{Universidad Complutense, Facultad de F\'{i}sicas, Ciudad Universitaria s/n, Madrid 28040, Spain}
\author{J.~I. Cirac}
\author{G. Rempe}
\affiliation{Max-Planck-Institut f{\"u}r Quantenoptik, Hans-Kopfermann-Stra{\ss}e 1, 85748 Garching, Germany}
\author{S. D\"{u}rr}
\affiliation{Max-Planck-Institut f{\"u}r Quantenoptik, Hans-Kopfermann-Stra{\ss}e 1, 85748 Garching, Germany}

\begin{abstract}
\normalsize
Atomic quantum gases in the strong--correlation regime offer unique possibilities to explore a variety of many--body quantum phenomena. Reaching this regime has usually required both strong elastic and weak inelastic interactions, as the latter produce losses. We show that strong inelastic collisions can actually inhibit particle losses and drive a system into a strongly--correlated regime. Studying the dynamics of ultracold molecules in an optical lattice confined to one dimension, we show that the particle loss rate is reduced by a factor of 10. Adding a lattice along the one dimension increases the reduction to a factor of 2000. Our results open up the possibility to observe exotic quantum many--body phenomena with systems that suffer from strong inelastic collisions.
\end{abstract}

\maketitle

Strong interactions are responsible for many interesting quantum phenomena in many--body systems: high-$T_C$ superconductivity \cite{anderson:87}, excitations with fractional statistic \cite{wilczek:82}, topological quantum computation \cite{kitaev:03}, and a plethora of exotic behaviors in magnetic systems \cite{auerbach:94}. One of the main physical mechanisms that gives rise to strong correlations for bosonic particles can be understood as follows. At low temperatures and for strong elastic repulsive interactions, particles tend to stay far away from each other in order to keep the energy low. That is, the wavefunction describing the particles tends to vanish when two of them coincide at the same position. In order to fulfill these constraints, this wavefunction has to be highly entangled at all times, which may give rise to counter--intuitive effects both in the equilibrium properties as well as in the dynamics. In 1D, for example, this occurs in the so--called Tonks--Girardeau gas (TGG) \cite{tonks:36,girardeau:60}, where the set of allowed wavefunctions for bosonic particles coincide (up to some transformation) with those of free fermions. Despite being bosons, the excitation spectrum, the evolution of the density distribution, etc, correspond to those of fermionic particles. In 2D the same mechanism leads to the fractional quantum Hall effect \cite{stormer:99}, where the ground state as well as the low energy excitations fulfill the above mentioned constraint, giving rise to the existence of anyons which behave neither like bosons nor fermions, but have fractional statistics \cite{wilczek:82}.

Here we show that inelastic interactions can be used to reach the strong correlation regime with bosonic particles: This may seem surprising because inelastic collisions are generally associated with particle losses. This behavior can be understood by using an analogy in classical optics, where light absorption is expressed by an imaginary part of the refractive index. If an electromagnetic wave impinges perpendicularly on a surface between two media with complex refractive indices $n_1$ and $n_2$, then a fraction $|(n_1-n_2)/(n_1+n_2)|^2$ of the intensity will be reflected. In the limit $|n_2|\rightarrow\infty$, the light is perfectly reflected off the surface, irrespective of whether $n_2$ is real or complex. In our case, bosons interacting with large imaginary \cite{verhaar:94,bohn:97} scattering length almost perfectly reflect off each other for an analogous reason, thereby giving rise to the same constraints in the particles' wavefunction as the ones corresponding to elastic collisions, and thus to the same physical phenomena. In our experiment, the correlations manifest themselves in a strong suppression of the rates at which particles are lost due to inelastic collisions.

Our experiment uses molecules confined to 1D by an optical lattice, both with and without a periodic potential along the 1D. We start with the transfer of a BEC of $^{87}$Rb atoms into a 3D optical lattice, in such a way that the central region of the resulting Mott insulator contains exactly two atoms at each lattice site. A Feshbach resonance at 1007.4 G \cite{marte:02} is used to associate the atom pairs to molecules \cite{duerr:04}. Subsequently, the magnetic field is held at 1005.5 G. Atoms remaining after the association are removed with blast light. This procedure prepares a quantum state that contains one molecule at each site of a 3D optical lattice \cite{volz:06,duerr:06:APS}. The optical--lattice potential seen by a molecule is $- V_\perp \cos^2(k x) - V_\perp \cos^2(k y) - V_\parallel \cos^2(k z)$, where $\lambda=2\pi/k=830.440$ nm is the light wavelength. At the end of the state preparation $V_\parallel=V_\perp=127 \, E_r$, where $E_r=\hbar^2 k^2/2m$ is the molecular recoil energy and $m$ is the mass of one molecule.

After state preparation, $V_\parallel$ is linearly ramped down to its final value. After this ramp, we have an array of tubes of 1D gases. We choose a ramp duration of 0.5 ms. For much faster ramps, we observe a substantial broadening of the momentum distribution along the tubes. For much slower ramps, particle loss during the ramp becomes noticeable. After the ramp, the system is allowed to evolve for a variable hold time at the final value of $V_\parallel$. During this hold time, molecules collide inelastically, leading to loss. After the hold time, all molecules are dissociated \cite{duerr:04} into atom pairs using the Feshbach resonance. The dissociation terminates the loss. Finally, the magnetic field and the lattice light are switched off simultaneously, and the number of atoms is determined from a time--of--flight absorption image.

\begin{figure}[tb!]
\includegraphics[width=0.95\columnwidth]{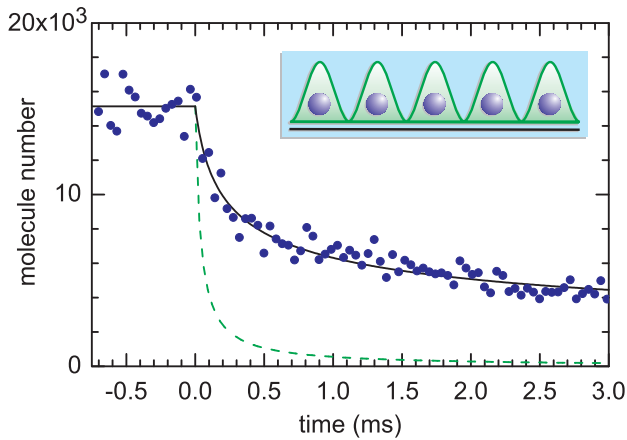}
\caption{
\label{fig-decay-curve} 
Time--resolved loss of molecules at $V_\parallel=0$. The loss begins at $t=0$. The solid line shows a fit of Eq.\ \ref{eq-n^4} to the experimental data ($\bullet$) with $t\leq 1$ ms. The best--fit value is $\chi n^3(0)=4.3/$ms, corresponding to $K_{3D}=2.2\times10^{-10}$ cm$^3$/s and, at $t=0$, to $g^{(2)}=0.11$. The dashed line shows the expectation for an uncorrelated system. The observed loss is much slower than the dashed line because of strong correlations.
}
\end{figure}

The experimental data represent an average over a large number of tubes of different lengths. This is not critical because the initial 1D density $n(0)=2/\lambda$ is identical in all tubes. The lattice beams that create $V_\perp$ have a finite waist. This results in a harmonic confinement with angular frequency $\omega_\parallel=2\pi\times 71$ Hz along the tubes. This is negligible as long as we evaluate the loss only for much faster timescales. In the decay of the molecule number as a function of hold time at $V_\parallel=0$ (Fig.\ \ref{fig-decay-curve}) the ramp down of $V_\parallel$ begins at $t=-0.5$ ms and ends at $t=0$. The data do not show noticeable loss during the ramp down. In order to avoid complications due to the harmonic confinement $\omega_\parallel$ along the tubes, we process only data for $t\leq 1$ ms.

A quantitative understanding of the loss process is based on the 1D particle density $n$, which evolves according to \cite{supplement}
\begin{eqnarray}
\label{eq-losses}
\frac{d n}{dt} = -K n^2 g^{(2)}
\; ,
\end{eqnarray}
where $g^{(2)}= \langle n^2 \rangle / \langle n \rangle^2$ is the pair correlation function that gives the reduction factor of the loss rate compared to an uncorrelated state where $g^{(2)}=1$. $K$ is the 1D loss rate coefficient, which can be related to the 3D scattering properties as follows: First, the scattering potential can be modeled as a delta interaction with 1D interaction strength $g$, yielding $K=-2 {\rm Im}(g)/\hbar$ \cite{supplement}. Second, extending the arguments of Ref.\ \cite{olshanii:98} to the case of a complex--valued 3D scattering length $a$, one obtains $g= 2\hbar^2 a/(m a_\perp^2 [ 1+ a\zeta\left(\frac12\right)/\sqrt2 \, a_\perp])$, where $a_\perp$ is the size of the gas in the perpendicular direction and $\zeta$ denotes the Riemann zeta function with $\zeta(\frac12)\approx -1.46$. The real and imaginary parts of $a$ represent elastic and inelastic scattering, respectively.

For the ground state of the TGG, one can derive an analytic expression for $g^{(2)}$. Introducing a dimensionless interaction strength $\gamma= m g/\hbar^2 n$ and using the same techniques as in Refs.\ \cite{lieb:63,gangardt:03} one obtains $g^{(2)}=4\pi^2/3|\gamma|^2$ in the limit $|\gamma|\gg 1$. Inserting this expression in Eq.\ \ref{eq-losses} we obtain
\begin{eqnarray}
\label{eq-n^4}
\frac{d n}{dt} = - \chi n^4
\; ,
\end{eqnarray}
where $\chi=4\pi^2\hbar^4 K / 3m^2|g|^2$. This equation gives a prediction for the particle losses for a low energy TGG. Similar considerations apply to inelastic three--body collisions \cite{kagan:85,burt:97,laburthe:04}. Note that a TGG at finite temperature has recently been observed experimentally \cite{paredes:04,kinoshita:04,kinoshita:05} but for the known case of strong elastic interactions.

A fit of Eq.\ \ref{eq-n^4} to the data in Fig.\ \ref{fig-decay-curve} with $t\leq 1$ ms yields ${\rm Im}(1/a)=1/(24 \ {\rm nm})$. Combined with $|{\rm Re}(a)| \ll |{\rm Im}(a)|$ \cite{supplement} this yields $g^{(2)}=0.11\pm 0.01$ at $t=0$. This shows that the loss rate is strongly reduced due to correlations. $g^{(2)}\propto n^2$ decreases with time, so that the loss rate is reduced even further.

\begin{figure}[tb!]
\includegraphics[width=0.95\columnwidth]{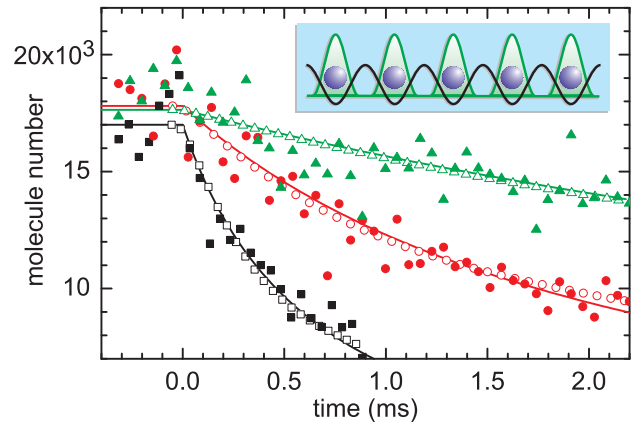}
\caption{
\label{fig-decay-lattice} 
Loss at $V_\parallel \neq 0$. Solid lines show fits of Eq.\ \ref{eq-loss-kappa} to the experimental data (filled symbols). Open symbols show results of our numerical calculations. Black squares, red circles, and green triangles correspond to $V_\parallel/E_r=1.8$, 3.9, and 6.0 respectively.
}
\end{figure}

From these results we can determine the two--body loss coefficient for a Bose--Einstein condensate $K_{3D}=-8\pi\hbar {\rm Im}(a)/m=(2.2\pm 0.2)\times10^{-10}$ cm$^3$/s. This agrees fairly well with the measurements discussed further below and with a previous measurement \cite{syassen:06} at 1005.8 G that yielded $K_{3D}=1.5\times 10^{-10}$ cm$^3$/s. For comparison we note, that if losses were not inhibited, {\it i.e.}, $g^{(2)}=1$, then the loss should follow the dashed line, which is calculated with $K_{3D}$ from Ref.\ \cite{syassen:06}. Clearly, the loss rate is reduced.

This reduction is directly related to the fact that the spatial wavefunctions of the particles do not overlap, and thus they become strongly correlated. Further support for this conclusion comes from the time dependence of our data. If the system were weakly correlated, then $g^{(2)}\approx 1$ would be time independent. Eq.\
\ref{eq-losses} would then predict that the number of particles $N$ follows $dN/dt \propto N^2$ instead of $dN/dt \propto N^4$ if we are close to the TGG ground state \cite{supplement}. We fit $dN/dt \propto N^p$ with an arbitrary power $p$ to the data with $t\leq1$ ms. This yields $p=4.3\pm 0.6$ in good agreement with $p=4$.

\begin{figure}[tb!]
\includegraphics[width=0.95\columnwidth]{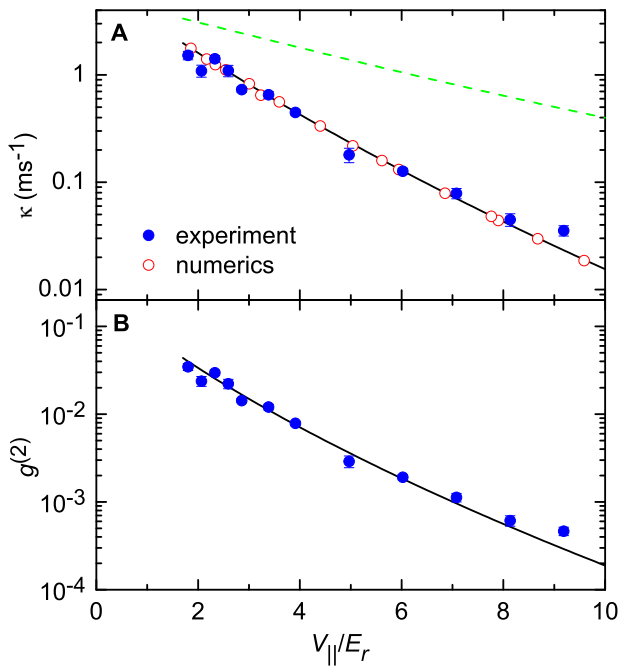}
\caption{\label{fig-kappa} Loss at different lattice depths $V_\parallel$. {\bf A} Fits as in Fig.\ \ref{fig-decay-lattice} yield the experimental results ($\bullet$) for $\kappa$. A fit of Eq.\ \ref{eq-kappa} to the data yields the solid line. The best--fit value is $K_{3D}=1.7\times10^{-10}$ cm$^3$/s. The experimental data and the analytic model agree well with results of our numerical calculations ($\circ$). For comparison, the dashed line shows the na\"{i}ve estimate $2J/\hbar$ which is nowhere even close to the data. {\bf B} Pair correlation function $g^{(2)}=\kappa/\Gamma$ calculated from the data in \bf A. }
\end{figure}

We now turn to a situation with a lattice potential along the 1D tubes with $V_\parallel \ll V_\perp$. The motion perpendicular to the tubes remains frozen out as before, but the motion along the tubes is now described as hopping between discrete lattice sites. This is a natural way of amplifying the effects due to interactions and thus reaching more deeply into the strong--correlation regime \cite{stoeferle:04,paredes:04}. Apart from that, many paradigmatic models in solid state and other fields of physics assume a lattice structure. We performed measurements (Fig.\ \ref{fig-decay-lattice}) similar to that in Fig.\ \ref{fig-decay-curve} for various values of the lattice depth $V_\parallel$. Here, the temporal change of $g^{(2)}$ is negligible \cite{supplement}. Spatial integration of Eq.\ \ref{eq-losses} yields 
\begin{eqnarray}
\label{eq-loss-kappa}
\frac{d N}{dt} = -\frac{\kappa}{N(0)} N^2(t)
\; ,
\end{eqnarray}
where we abbreviated $\kappa = K n(0) g^{(2)}$. We fit this to the data with $N(t)\geq N(0)/2$ in order to neglect the harmonic confinement along the tube. The best--fit values for the loss rate $\kappa$ obtained for different lattice depths $V_\parallel$ are shown in Fig.\ \ref{fig-kappa}A. 

\begin{figure}[tb!]
\includegraphics[width=0.6\columnwidth]{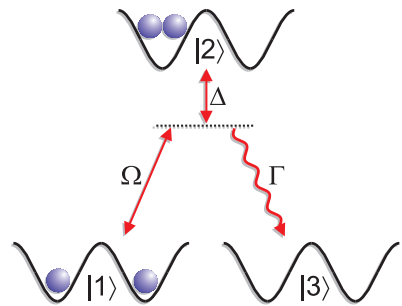}
\caption{
\label{fig-scheme}
Particle losses in a 1D lattice. The initial level $|1\rangle$ contains exactly one particle at each lattice site. State $|2\rangle$ is obtained after one tunneling process. Population in state $|2\rangle$ decays incoherently into state $|3\rangle$ with rate $\Gamma$. The tunneling coupling between states $|1\rangle$ and $|2\rangle$ can be described by a Rabi frequency $\Omega$ and a detuning $\Delta$. In the limit $\Gamma\gg\Omega$ the effective decay rate from $|1\rangle\to|3\rangle$ is $\Gamma_{\rm eff}=\Omega^2/(\Gamma [1+(2\Delta/\Gamma)^2])$ and the population of level $|2\rangle$ after a time $t\gg 1/\Gamma$ is $\Gamma_{\rm eff}/\Gamma$ times that of level $|1\rangle$ \cite{cohen-tannoudji:92:p49}. The parameters are related to the Bose--Hubbard parameters by $\Gamma=-2{\rm Im}(U)/\hbar$, $\Delta={\rm Re}(U)/\hbar$, and $\Omega=\sqrt{8}J/\hbar$.
}
\end{figure}

A prediction for $\kappa$ can be developed starting from the Lieb--Liniger Hamiltonian \cite{lieb:63,supplement}. In the presence of a periodic potential in the tight--binding limit, one obtains a Bose--Hubbard model \cite{jaksch:98} with tunneling amplitude $J$ and on--site interaction ${\rm Re}(U)$. $J$ and $U$ can be expressed in terms of $m$, $a$, and a Wannier function. $\Gamma=-2{\rm Im}(U)/\hbar$ is the rate at which two particles at the same site are lost. In this model losses occur when two particles occupy neighboring sites and one of them hops to the other's site. For strong inelastic interactions ($J/\hbar\Gamma\ll 1$) one obtains an effective loss rate $\Gamma_{\rm eff}$ for two neighboring particles (Fig.\ \ref{fig-scheme}). An extension of this double--well model to many sites yields $\kappa=4 \Gamma_{\rm eff}$ \cite{supplement} so that
\begin{eqnarray}
\label{eq-kappa}
\kappa = \frac{32J^2}{\hbar^2\Gamma} \left[ 1+ \left(\frac{{\rm Re}(a)}{{\rm Im}(a)}\right)^2 \right]^{-1}
\; .
\end{eqnarray}
This coefficient shows again that for strong inelastic interactions ($\Gamma \to\infty$) particle losses are inhibited. For large, finite values of $\Gamma$ one can, as before, observe such an inhibition by looking at the particle losses. We fit Eq.\ \ref{eq-kappa} to the data in Fig.\ \ref{fig-kappa}A. With $|{\rm Re}(a)| \ll |{\rm Im}(a)|$ there is only one free fit parameter. The best--fit value is $K_{3D}=(1.7\pm 0.3)\times10^{-10}$ cm$^3$/s, which is close to the result of Fig.\ \ref{fig-decay-curve}. As $V_\parallel/E_r$ increases from 1.7 to 10, $\Gamma$ increases from 45/ms to 82/ms.

In order to test the quality of this analytic model, we performed extensive numerical calculations by solving the master equation \cite{supplement} using matrix product density operators \cite{verstraete:04}. They reveal that the system maps to a very good approximation to a fermionized gas and that it loses its memory about the initial state in a time $\sim 1/\Gamma$. During this very short time only little loss occurs, and after this short transient the loss is well described by a time independent $g^{(2)}$ with $\kappa$ from Eq.\ \ref{eq-kappa}. We find good agreement between experimental data, analytic model, and numerical results in Figs.\ \ref{fig-decay-lattice} and \ref{fig-kappa}.

For $|{\rm Re}(a)| \ll |{\rm Im}(a)|$, Eq.\ \ref{eq-kappa} tells us that the larger the loss coefficient $\Gamma$, the smaller the actual loss rate $\kappa$. This means that fast on--site loss tends to preserve the initial state and thus suppresses tunneling in the many--body system. This can be interpreted as a manifestation of the continuous quantum Zeno effect \cite{misra:77}: fast dissipation freezes the system in its initial state. Without this Zeno effect, one might na\"{i}vely estimate that tunneling would occur at a rate $\sim 2J/\hbar$. If each such tunneling event would lead to immediate loss, then $2J/\hbar$ should set the timescale for the loss, but that estimate is too na\"{i}ve, as the dashed line in Fig.\ \ref{fig-kappa}A shows.

The value of $K_{3D}$ extracted from Fig.\ \ref{fig-kappa}A is used to calculate $\Gamma$ and thus $g^{(2)}=\kappa/\Gamma$ \cite{supplement} for each experimental data point. The results (Fig.\ \ref{fig-kappa}B) agree well with the theoretical expectation (solid line) based on the same value of $K_{3D}$. The smallest measured value of $g^{(2)}=(4.6\pm 0.7)\times 10^{-4}$ represents an improvement of more than two orders of magnitude over previous experiments \cite{kinoshita:05}. Note that a noninteracting gas has $g^{(2)}=1$ at any lattice depth. The observed suppression of $g^{(2)}$ is caused by the inelastic interactions, not by the lattice itself. Moreover, the strength of the correlations in our experiment is determined by the interparticle interactions. This differs fundamentally from experiments in very deep lattices where tunneling is negligible on the timescale of the experiment, interactions are irrelevant, and bosonic symmetrization of the wave function is possible but has no detectable consequences.

The mechanism introduced here could also give rise to other strongly--correlated states, such as a Laughlin state \cite{stormer:99} or one with anyonic excitation \cite{wilczek:82}. The present paper opens up the possibility of observing exotic quantum many--body phenomena in systems that suffer from strong inelastic collisions. Furthermore, the rate coefficients for those collisions may be artificially increased using photoassociation or Feshbach resonances, thus further reducing the actual loss rate in the strongly--correlated regime.

\clearpage

\renewcommand{\theequation}{\mbox{S\arabic{equation}}}
\setcounter{equation}{0}

\noindent
\begin{minipage}{\textwidth}
\begin{center} 
{\large Supplementary Online Material for \\[1ex]}
{\bf \large Strong Dissipation Inhibits Losses and Induces Correlations in Cold Molecular Gases \\[1ex] }
N. Syassen, D.~M. Bauer, M. Lettner, T. Volz, D. Dietze, J.~J. Garc\'{i}a-Ripoll, J.~I. Cirac, G. Rempe, and S. D\"{u}rr \end{center}
\end{minipage}

\section*{MODELING DETAILS}

\section*{Real Part of the Scattering Length}
First, we analyze the contributions of ${\rm Re}(a)$ and ${\rm Im}(a)$ to the value ${\rm Im}(1/a)=1/(24~\rm nm)$ extracted from Fig.\ \ref{fig-decay-curve}. The identity
\begin{eqnarray}
\label{eq-sup-Re-a^2}
{\rm Re}(a)^2 = - {\rm Im}(a)^2 - \frac{{\rm Im}(a)} {{\rm Im}(1/a)}
\end{eqnarray}
shows that ${\rm Re}(a)^2$ is a parabola as a function of ${\rm Im}(a)$ at fixed ${\rm Im}(1/a)$. This yields an upper bound $|{\rm Re}(a)|\leq 1/(2 {\rm Im}(1/a))=12$ nm. This information can be combined with our previous studies of excitation spectra \cite{volz:06,duerr:06:APS}. With the upper bound for $|{\rm Re}(a)|$, the excitation spectra should yield a resonance at a frequency below 1.4 kHz for $V_\parallel=V_\perp=15\, E_r$. We performed measurements that cover a broad range of experimental parameters but could not detect such a resonance. This can only be explained if $|{\rm Re}(a)| \ll |{\rm Im}(a)|$, because in this case the resonance becomes very broad and shallow, so that it might be undetectable. We conclude that $|{\rm Re}(a)| \ll |{\rm Im}(a)|$. As a result ${\rm Im}(1/a) \approx - 1/{\rm Im}(a)$.

The 1D measurements (no matter if $V_\parallel=0$ or $V_\parallel\neq0$) yield an experimental value for ${\rm Im}(1/a)$, while the 3D measurement in Ref.\ \cite{syassen:06} yields a value for ${\rm Im}(a)$. In principle, these two values could be inserted into Eq.\ \ref{eq-sup-Re-a^2} to extract a value (not just an upper bound) for $|{\rm Re}(a)|$. But in practice the systematic uncertainty in Ref.\ \cite{syassen:06} is so large, that we cannot constrain $|{\rm Re}(a)|$ any further.

\section*{Dissipative Lieb--Liniger Model}
Now we turn to modeling the loss in the 1D tubes at $V_\parallel=0$. A 1D spinless bosonic system in a box with periodic boundary conditions (similar conclusions apply to many other situations) is described by the Lieb--Liniger Hamiltonian \cite{lieb:63}
\begin{eqnarray}
\label{SOM-eq-H}
H = - \frac{\hbar^2}{2m} \sum_i \frac{\partial^2}{\partial x_i^2} + g \sum_{i<j} \delta (x_i-x_j)
\; ,
\end{eqnarray}
where $x_i$ is the position of the $i$th boson. Generalizing the techniques developed in Ref.\ \cite{lieb:63} to the case of complex--valued $g$, one can show that the imaginary parts \linebreak[4]

\vspace*{25mm}
\noindent
of all eigenvalues of $H$ vanish as $|g|\to\infty$ and that the corresponding eigenfunctions no longer overlap. The former indicates that in this limit there will be no losses, while the latter gives rise to strong--correlation phenomena. For small values of the relative position of two particles, $x_{ij}=x_i-x_j$, the eigenfunctions have to scale as $\psi \propto 2\hbar^2/m g + |x_{ij}| +{\cal O}(x_{ij}^2)$ in order to satisfy the Schr\"{o}dinger equation. In the above limit and for $x_{ij}=0$, $\psi$ must vanish regardless of the argument of $g$. In any experimental situation, $|g|$ will be finite. In that case, there will be finite losses but they will remain small if $|g|$ is large. The master equation discussed further below can be used to derive the equation describing the evolution of the 1D particle density $n$
\begin{eqnarray}
\label{SOM-eq-losses}
\frac{d \langle n\rangle}{dt}=-K g^{(2)} \langle n\rangle^2 
\; ,
\end{eqnarray}
where $K=-2 {\rm Im}(g)/\hbar$ is a 1D loss rate coefficient and $g^{(2)}= \langle n^2 \rangle / \langle n \rangle^2$ is the pair correlation function. Time integration of Eq.\ \ref{eq-n^4} followed by spatial integration yields $N(t)=N(0)(1+3n^3(0)\chi t)^{-1/3}$.

\section*{Dissipative Lattice Model}
We now turn to a situation with a lattice along the 1D tubes and with periodic boundary conditions. The time evolution of the particle number $n_k$ at site $k$ can be derived from the master equation discussed further below. The result closely resembles Eq.\ \ref{SOM-eq-losses} 
\begin{eqnarray}
\label{SOM-eq-loss-nk}
\frac{d \langle n_k\rangle}{dt}=-\Gamma g^{(2)} \langle n_k\rangle ^2 
\; ,
\end{eqnarray}
where in the lattice $g^{(2)}=\langle n_k(n_k-1)\rangle/\langle n_k\rangle^2$. In order to obtain an analytic approximation for $g^{(2)}$ we consider the limit $J\ll \hbar\Gamma$, where the probability to have more than two particles at one site is negligible. Hence $g^{(2)}=2p_2/p_1^2$ where $p_i$ is the probability of having $i$ particles at site $k$. We then use the three--level model of Fig.\ \ref{fig-scheme}. We can write $p_2=q_{11}\Gamma_{\rm eff}/\Gamma$, where $q_{11}$ is the probability that both site $k$ and one of its neighbors is occupied, which we can estimate as $q_{11}\sim 2p_1^2$. This approximation is motivated by the observation that for $J\ll\hbar\Gamma$ the three--level model of Fig.\ \ref{fig-scheme} yields $\hbar \Gamma_{\rm eff} \ll J$. Hence, the particles have time to redistribute across the lattice between successive loss events. Our numerical results confirm that this is a good approximation in the parameter regime of our experiment. Thus we get
\begin{eqnarray}
g^{(2)}=\kappa/\Gamma
\end{eqnarray}
with $\kappa=4\Gamma_{\rm eff}$. The approximation $q_{11}\sim 2p_1^2$ makes $g^{(2)}$ independent of density and therefore independent of time, in contrast to the case $V_\parallel=0$. $g^{(2)}$ can be made extremely small by reducing the tunneling amplitude, and thus one can approach a fermionized gas. Furthermore, assuming that each lattice site is initially occupied by one particle, spatial integration of Eq.\ \ref{SOM-eq-loss-nk} yields
\begin{eqnarray}
\label{SOM-eq-loss-kappa}
\frac{d N}{dt} = -\frac{\kappa}{N(0)} N^2(t)
\; .
\end{eqnarray}
Time integration yields $N(t)=N(0)/(1+\kappa t)$.

\section*{Master Equation}
In general, particle losses are described in terms of a master equation for the density operator $\rho$ describing the particles
\begin{eqnarray}
\hbar \frac{d\rho}{dt}= -i H\rho + i\rho H^\dagger + {\cal J}(\rho)
\; .
\end{eqnarray}
For the case without the lattice along the 1D tubes, $H$ is the second--quantized version of Eq.\ \ref{SOM-eq-H} and
\begin{eqnarray}
{\cal J}(\rho)= - {\rm Im}(g) \int dx \hat \Psi^2 (x) \rho \hat \Psi^{\dagger 2} (x)
\; ,
\end{eqnarray}
where $\hat \Psi(x)$ is the bosonic field operator. If we assume $|{\rm Im}(g)|\to\infty$, then the wavefunctions do not overlap and thus ${\cal J}(\rho)=0$. Thus, the master equation is reduced to a simple Schr\"{o}dinger equation (for an initially pure state), $\hbar d|\Psi\rangle/dt= -i H|\Psi\rangle$, where we can make use of the fermionization transformation of Girardeau. On the other hand, for finite losses, we can make use of the master equation to derive the evolution of the total particle number
\begin{eqnarray}
N=\int dx \langle\Psi^{\dagger} (x) \hat \Psi (x)\rangle
\; ,
\end{eqnarray}
obtaining Eq.\ \ref{SOM-eq-losses} for a homogeneous system.

In the presence of the lattice along the 1D tubes, we obtain a Bose--Hubbard model
\begin{eqnarray}
H &=& -J\sum_k (a^\dagger_k a_{k+1} + H.c.) + \frac{U}{2}\sum_k a^{\dagger 2}_k a_k^2 \qquad
\\
{\cal J}(\rho)&=& - {\rm Im}(U) \sum_k a_k^2 \rho  a_k^{\dagger 2}
\end{eqnarray}
and the same as before can be done.

Our numerical calculations are based on this master equation. As $J \ll \hbar\Gamma$, they neglect the possibility that more than two particles could occupy the same site. The lattice depth is modeled as time--independent and the evolution of the particle number starting from a quantum state with one particle at each lattice site is modeled without any assumptions about the time dependence of $g^{(2)}$.

\section*{Finite Energy in the Experiment}
Finally, we address the question, how close our experiment is to the ground state of the TGG. Eq.\ \ref{eq-n^4} assumes that we have a TGG at low energy, close to the ground state. Thus, the reasonable agreement of our experimental data with that equation gives a strong indication that indeed we are relatively close to the ground state. As the loss proceeds, the particle number changes and the ground state changes correspondingly. A simple estimate for the temporal evolution of the energy can be obtained when assuming that the kinetic energy per particle $E^{\rm kin}/N$ would remain constant when two particles are lost. This is to be compared to the new energy of the ground state which now has two less particles. In a harmonic oscillator, the kinetic energy of the fermionized ground state scales as $E_g^{\rm kin} \propto N^2$. As the loss proceeds, the system thus automatically evolves away from the ground state, because $E^{\rm kin}(t)/E_g^{\rm kin}(t)=N(0)/N(t)$. Actually, this gives us an upper bound to the energy. The reason is that in the limit studied here, for the (pseudo) eigenstates of Eq.\ \ref{SOM-eq-H} the imaginary part of the corresponding eigenvalues is proportional to the real (kinetic) energy. Thus, the states with higher kinetic energies are preferentially depleted. In any case, ignoring this last fact we estimate for the data with $t\leq 1$ ms in Fig.\ \ref{fig-decay-curve} that $E^{\rm kin}/E_g^{\rm kin}$ increases by a factor of $\sim 2$ during the loss. This is comparable to the energy of the TGG observed in Ref.\ \cite{paredes:04}, where temperatures of the order of the Fermi energy were achieved.

\clearpage

\end{document}